\documentstyle[twoside]{article} \oddsidemargin 0cm \evensidemargin 0cm
\def\Red{} \def\Black{} \def\Blue{}

\renewcommand{\Im}{\mathop{\rm Im}} \newcommand{\ecm}{\,e\cdot{\rm cm}}
\newcommand{\GeV}{\,{\rm GeV}} \newcommand{\TeV}{\,{\rm TeV}}
 \newcommand{\NP}[3]{{\em Nucl. Phys. \bf #1}
  (#2) #3} \newcommand{\PRL}[3]{{\em Phys. Rev. Lett. \bf #1} (#2) #3}
\newcommand{\PL}[3]{{\em Phys. Lett. \bf #1} (#2) #3}

\makeatletter
\newcounter{alphaequation}[equation]
\def\thealphaequation{\theequation\hbox to
0.6em{\hfil\alph{alphaequation}\hfil}}
\def\eqnsystem#1{
\def\@eqnnum{{\rm (\thealphaequation)}}
\def\@@eqncr{\let\@tempa\relax \ifcase\@eqcnt \def\@tempa{& & &} \or
  \def\@tempa{& &}\or \def\@tempa{&}\fi\@tempa
  \if@eqnsw\@eqnnum\refstepcounter{alphaequation}\fi
\global\@eqnswtrue\global\@eqcnt=0\cr}
\refstepcounter{equation} \let\@currentlabel\theequation \def\@tempb{#1}
\ifx\@tempb\empty\else\label{#1}\fi
\refstepcounter{alphaequation}
\let\@currentlabel\thealphaequation
\global\@eqnswtrue\global\@eqcnt=0 \tabskip\@centering\let\\=\@eqncr
$$\halign to \displaywidth\bgroup \@eqnsel\hskip\@centering
$\displaystyle\tabskip\z@{##}$&\global\@eqcnt\@ne
\hskip2\arraycolsep\hfil${##}$\hfil& \global\@eqcnt\tw@\hskip2\arraycolsep
$\displaystyle\tabskip\z@{##}$\hfil
\tabskip\@centering&\llap{##}\tabskip\z@\cr}

\def\endeqnsystem{\@@eqncr\egroup$$\global\@ignoretrue} \makeatother

  \def\Ord{{\cal O}}  \def\SU{{\rm SU}}  
\def\circa#1{\,\raise.3ex\hbox{$#1$\kern-.75em\lower1ex\hbox{$\sim$}}\,}

\begin{document}
November 1995\hfill\vbox{ \hbox{\bf IFUP -- TH--65--95}\hbox{\bf
    hep-ph/9511305}}\\[5mm]

\centerline{\huge\bf\Red Electric dipole moments as} \centerline{\huge\bf
  signals of supersymmetric unification}\bigskip\bigskip\Black
\centerline{\large\bf Riccardo Barbieri, Andrea Romanino {\rm and} Alessandro
  Strumia} \bigskip \centerline{\large\em Dipartimento di Fisica, Universit\`a
  di Pisa} \centerline{\large and} \centerline{\large\em INFN, Sezione di Pisa,
  I-56126 Pisa, Italy} \bigskip\bigskip\Blue \centerline{\large\bf Abstract}
\begin{quote}\large\indent
  If supersymmetric unification is true, we show how the combined effort of
  several experiments under way to try to measure an electric dipole moment of
  the electron or of the neutron has a significant chance not only of producing
  a positive signal but also of providing crucial information to understand the
  physical origin of the signal itself.
\end{quote}\Black

\section{}\label{Introduction}
As recently pointed out~\cite{DH, 9Art}, the electric dipole moments (EDMs) of
the electron, $d_e$, and of the neutron, $d_N$, represent a very significant
signature for supersymmetric unification.  In a typical supersymmetric GUT with
supersymmetry breaking transmitted by supergravity couplings, the heaviness of
the top quark induces a splitting between the sfermion masses of the third
generation with respect to the masses of the first two
generations~\cite{BH,8Art}.  Such splitting, together with the CKM-like mixing
angles and phases appearing in the gaugino-matter interactions, manifests
itself, through one loop radiative corrections, in electron and neutron EDMs
which are at the level of the current limits for large CP-violating phases and
for sparticle masses visible at LHC~\cite{DH, 9Art}.  This observation
justifies the believe that the electron and neutron EDMs can be considered
among the few characteristic signatures of supersymmetric unification and
should therefore be vigorously searched for.  This view is strengthened by the
fact that the discovery of the EDMs, if indeed originated by the unified
interactions, must be accompanied by the observation of processes with
violation of lepton flavour, such as $\mu\to e\gamma$ or $\mu\to e$ conversion
in atoms, with typical rates again ``around the corner''~\cite{BH,8Art}.

Needless to say, however, as always in the case of radiative corrections
effects, the discovery of an EDM would not allow an immediate identification of
its physical origin.  In general one would have to discriminate between sources
of EDMs inside or beyond the Standard Model (SM) or even, within a definite
extension of the SM, between alternative mechanisms that can produce an EDM.

This last case is of relevance to the theories of interest to this paper.  In a
generic supersymmetric extension of the SM with minimal particle content one
can identify four different sources of CP violation and, eventually, of EDMs:
\begin{itemize}
\item[i.] the CKM phase in the charged current interactions;
\item[ii.] the strong $\theta_{\rm QCD}$-angle;
\item[iii.] the phases appearing in the soft terms of the supersymmetry
  breaking Lagrangian (``complex soft terms'' case);
\item[iv.] other CKM-like phases entering the fermion-sfermion-gaugino
  (higgsino) interactions (``unified theory'' case).
\end{itemize}
The first two sources are in common with the SM; the third one might be present
in any softly broken supersymmetric Lagrangian; the last one is present, at a
significant level, in unified theories like SO(10).  It is therefore at least
the last case that one wants to discriminate against the others.  We intend to
show under which circumstances this might be possible: special attention must
be payed to compare the results expected in cases iii.\ versus iv.

\section{}
On the experimental side, the search for EDMs is being pursued~\cite{Ramsey} by
working on different systems: the paramagnetic atoms with open shells of
unpaired electron spins, the diamagnetic atoms with closed shells of paired
electron spins, and the neutron.  As it will be immediately clear, this
diversity of experimental searches is essential.

{}From a microscopic point of view~\cite{Thomas}, or more precisely in terms of
the physics at the Fermi scale, all the aforementioned sources of CP violation
can affect the experiments on the EDMs in a significant way only through the
electron EDM, $d_e$, the EDM of the up quark, $d_u$ and of the down quark,
$d_d$, generically denoted by $d_q$, the chromoelectric dipole moments of the
same quarks, $d_q^{\rm QCD}$, and the $G_{\mu\nu}\tilde{G}^{\mu\nu}$ operator
in the QCD Lagrangian.  Other contributions from the three-gluon
operator~\cite{Weinberg} or from four-fermion interactions do not play any
significant role in the present discussion.  It is on the other hand well
known~\cite{Thomas} that the three kinds of experiments considered are affected
in a different way by the electron EDM and by the quark dipole moments,
electric or chromoelectric.

Taking three among the most significant cases, one per category, one has for
the respective EDMs~\cite{Thomas}
\begin{eqnsystem}{sys:dipoles}
\parbox{8em}{paramagnetic:}&
d_{\rm Tl}\!\! &=
\parbox{3em}{$-600 d_e$} +
\parbox{5em}{$\Ord (10^{-4}) d_q$} +
\parbox{6em}{$\Ord (10^{-3})d_q^{\rm QCD}$} +
\Ord (10^{-3}) (\theta /10^{-9}) d_{\rm Tl}^{1995}\\
\parbox{8em}{diamagnetic:}&
d_{\rm Xe}\!\! &=
\parbox{3em}{$10^{-3} d_e$}+
\parbox{5em}{$\Ord(10^{-4})d_q$} +
\parbox{6em}{$\Ord(10^{-3})d_q^{\rm QCD}$}
+ \Ord (10^{-1})(\theta /10^{-9}) d_{\rm Xe}^{1995}\\
\parbox{8em}{neutron:}&
d_N \!\! &=
\parbox{4.25em}{~}\makebox[0em][r]{1.6}
\parbox{5em}{$(\frac{4}{3}d_d-\frac{1}{3}d_u)$}+
\parbox{6em}{${\cal O}(10^{-1}) d_q^{\rm QCD}$}+
\Ord(1) (\theta /10^{-9}) d_N^{1995}
\end{eqnsystem}
where we have expressed the contribution from strong CP-violation involving the
$\theta_{\rm QCD}$ parameter in terms of the current upper bounds~\cite{Tl, Xe,
  N}
\begin{equation}
\label{eq:limiti}
d_{\rm Tl}^{1995} = 6.6\cdot 10^{-24}\ecm,\qquad d_{\rm Xe}^{1995} = 1.4\cdot
10^{-26}\ecm,\qquad d_N^{1995} = 0.8\cdot 10^{-25}\ecm
\end{equation}
respectively.

Let us assume possible improvements in the sensitivities of the various
experiments by one or two orders of magnitude at most~\cite{Ramsey}.  This
excludes the detection of an EDM generated by the CKM phase.  On the other
hand, in view of eq.s~(\ref{sys:dipoles}), (\ref{eq:limiti}) and of the range
of values taken by $d_e$, $d_q$, $d_q^{\rm QCD}$ in the theories under
consideration,
\begin{itemize}
\item[i.] the measurement of $d_{\rm Tl}$, or of the EDM for other paramagnetic
  systems, can be viewed as a search for an electron EDM;
\item[ii.] the measurement of $d_N$ might reveal a quark EDM or a strong
  CP-violation effect;
\item[iii.] the EDM of a diamagnetic system, like the Xe atom, might be
  influenced by all the three sources of CP-violation, from ii.~to iv., listed
  before.
\end{itemize}
In view of these considerations, we focus on the correlations between $d_e$,
which can be considered as a direct observable, and $d_N$.

\begin{figure}[t]\setlength{\unitlength}{1in}
\begin{center}\begin{picture}(8,3.5)(0.5,0)
  \put(0,0){\includegraphics{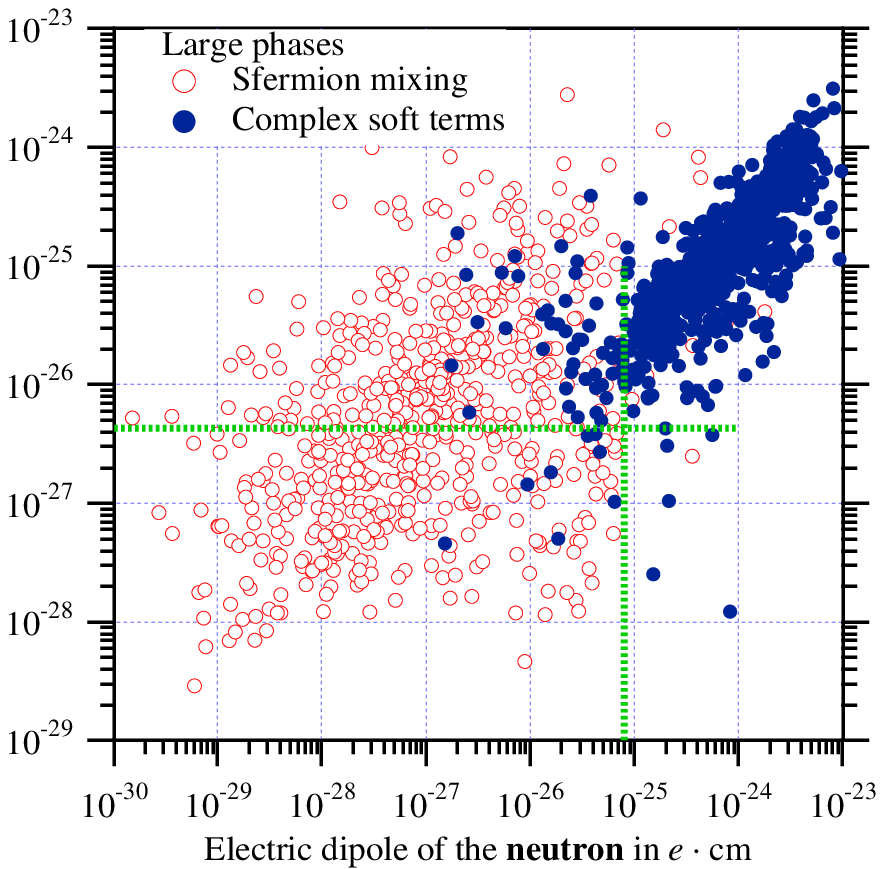}}
\put(3.7,0){\includegraphics{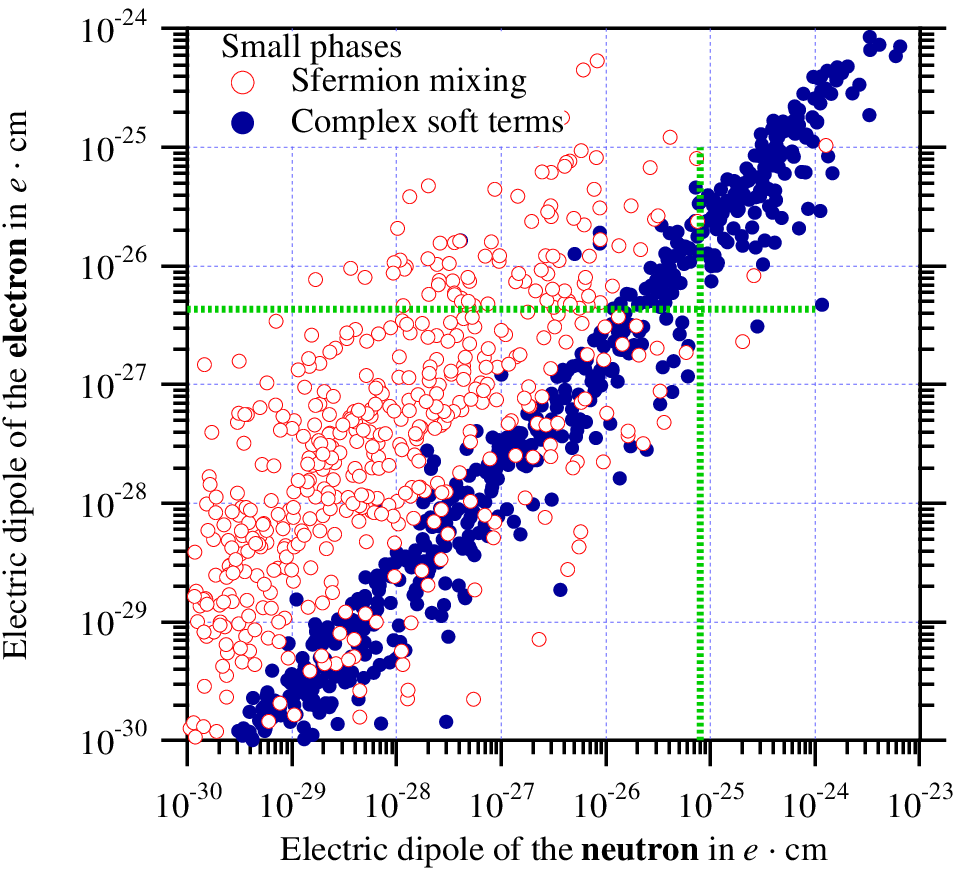}}
  \put(1.8,3.5){Fig.~\ref{fig:dnde}a} \put(5.6,3.5){Fig.~\ref{fig:dnde}b}
\end{picture}
\caption[1]{Scatter plot of for $d_N$ and $d_e$ as generated from
  complex soft terms ($\bullet$) and fermion-sfermion mixing matrices ($\circ$)
  in the cases of large (fig.\ \ref{fig:dnde}a) and small (fig.\
  \ref{fig:dnde}b) CP-violating phases (see the text for the range of the other
  parameters).  The dotted lines denote the present experimental upper bounds.
\label{fig:dnde}}
\end{center}\end{figure}

\section{}
Since the contribution to $d_N$ from strong CP-violation, for given
$\theta_{\rm QCD}$, is known, within a decent approximation, we concentrate on
$d_e$ and $d_N$ as arising from CP-violating phases in the soft supersymmetry
breaking terms (source iii.\ of section~\ref{Introduction}) or from CKM-like
phases in loops of sfermions and gauginos-higgsinos (source iv.\ of
section~\ref{Introduction}). More precisely, in connection with case iii., we
consider the MSSM with complex soft terms and universal initial conditions at
the GUT scale.  In this case, it is well known that only two phases have a
physical meaning.  We choose them to be the phase of the universal $A$-term and
of the $\mu$-parameter
\begin{equation} \label{eq:phases}
  A = |A|e^{i\phi_A},\qquad \mu = |\mu|e^{-i\phi_B},\qquad \hbox{with}\qquad
  B\mu = |B\mu|.
\end{equation}
As a prototype example of case iv., we consider the ``minimal'' SO(10)
theory~\cite{DH, 9Art, 8Art} with no other phases than in the Yukawa couplings
and with universal initial conditions on the soft terms at the Planck scale.

Both the electron and the quark EDMs, as the chromoelectric dipole moments, are
produced by one loop vertex diagrams with sfermions and gaugino-higgsinos as
internal lines.  In turn, the calculation of such diagrams involves the
knowledge of the full Lagrangian at the Fermi scale, which is the relevant
scale.  How the MSSM parameters are renormalized from their initial conditions
at $M_{\rm G}$ is too well known to be recalled here. In the precise case of
``minimal'' SO(10) with a large top Yukawa coupling, the rescaling to low
energy of the various parameters is done in ref.~\cite{8Art}.

In terms of these parameters, the various EDMs are readily computed by means of
the following formul\ae.  In the ``complex soft terms'' case one has
\begin{eqnsystem}{sys:soft}
  d_e &=& +\frac{e\alpha_{\rm e.m}}{4\pi\cos^2\theta_{\rm W}} m_e\Im
  (A_e^*+\mu\tan\beta)\sum_{n=1}^4\nonumber \frac{H_{n\tilde{B}}}{M_{N_n}}
  (H_{n\tilde{B}}+H_{n\tilde{W}_3}\cot\theta_{\rm W})
  G_2(\tilde{e}_L,\tilde{e}_R)+\\ &&+\frac{e\alpha_{\rm
      e.m.}}{4\pi\cos^2\theta_{\rm W}}\nonumber
  \frac{m_e}{\cos\beta\sin\theta_{\rm W}}\Im\sum_{n=1}^4
  \frac{H_{n\tilde{h}_{\rm d}}}{M_Z M_{N_n}}\left[ H_{n\tilde{B}}
  g_2(\frac{m^2_{\tilde{e}_R}}{M^2_{N_n}}) -\frac{1}{2}
  (H_{n\tilde{B}}+H_{n\tilde{W}_3}\cot\theta_{\rm W})
  g_2(\frac{m^2_{\tilde{e}_L}}{M^2_{N_n}})\right]+\\ &&+\frac{e\alpha_{\rm
      e.m.}}{4\pi\sin^2\theta_{\rm W}} \frac{m_e}{\cos\beta\cos\theta_{\rm
      W}}\Im \sum_{i=1}^2\frac{H^+_{i\tilde{W}} H^-_{i\tilde{h}_{\rm d}^-}}
{M_Z M_{\chi_i}} h_2(\frac{m_{\tilde{\nu}_L}^2}{M_{\chi_i}^2})\\
d_u &=& -\frac{16}{9}\frac{e\alpha_3}{4\pi M_3}m_u\Im[A_u^*+\mu\cot\beta]
G_2(\tilde{u}_L,\tilde{u}_R,3)\\ d_d &=& +\frac{8}{9}\frac{e\alpha_3}{4\pi
  M_3}m_d\Im[A_d^*+\mu\tan\beta]
G_2(\tilde{d}_L,\tilde{d}_R,3)\\
d_u^{\rm QCD} &=& +\frac{3}{2} \frac{g_3\alpha_3}{4\pi
  M_3}m_u\Im[A_u^*+\mu\cot\beta]
[H_2(\tilde{u}_L,\tilde{u}_R,3)+{2\over9}G_2(\tilde{u}_L,\tilde{u}_R,3)]\\
d_d^{\rm QCD} &=& +\frac{3}{2} \frac{g_3\alpha_3}{4\pi
  M_3}m_d\Im[A_d^*+\mu\tan\beta]
[H_2(\tilde{d}_L,\tilde{d}_R,3)+{2\over9}G_2(\tilde{d}_L,\tilde{d}_R,3)]
\end{eqnsystem}
where $H$, $H^+$, $H^-$ are the mass-eigenstate interaction-eigenstate rotation
matrices for the neutral $N_n$, the positively charged and the negatively
charged $\chi_i$ gauginos-higgsinos respectively, and
\begin{eqnsystem}{sys:def}
  g_2(r) &=& {1\over2(r-1)^3}[r^2-1-2r\ln r],\qquad h_2(r) =
  \frac{1}{2}-g_2(\frac{1}{r})\\ G_2(a,b,n) &=&
  \frac{g_2(m^2_a/M^2_n)-g_2(m^2_b/M^2_n)}{m^2_a-m^2_b} \\ H_2(a,b,n) &=&
  \frac{h_2(m^2_a/M^2_n)-h_2(m^2_b/M^2_n)}{m^2_a-m^2_b}.
\end{eqnsystem}
In the ``unified theory'' case the electric dipoles have already been computed
in~\cite{9Art,8Art}, to which we refer.

The relationship between these contributions is the following. In the ``complex
soft terms'' case all the contributions to $d_e$ not proportional to the
combination $(A_e^* + \mu\tan\beta)$ originate from the phase in the
gaugino-higgsino mass matrices, through the $\mu$-term. These terms, which
actually dominate the overall $d_e$, are of course not present in the ``unified
theory'' case. In such case, since all effects arise from flavour mixings in
the gaugino-matter interactions, they are subject to a potential GIM-like
cancellation.  Finally, at least in the ``minimal'' SO(10) case considered
here, there is no contribution to the up-quark DMs, either electric or
chromoelectric, because the mass matrices for the $Q=2/3$ quarks and squarks
may be simultaneously diagonalized without introducing any relative rotation.

\begin{figure}[t]\setlength{\unitlength}{1cm}
\begin{center}
\begin{picture}(10,9)
  \put(-1,0){\includegraphics{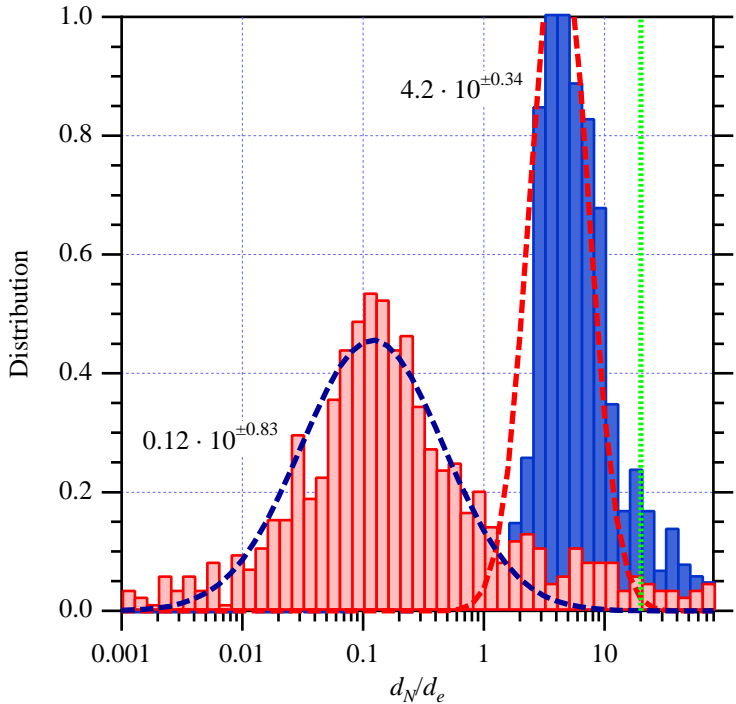}}
\end{picture}\hspace{0.3cm}\raisebox{4cm}{
\parbox[b]{5cm}{\caption{Distribution of the values of $d_N/d_e$ as generated
from
    complex soft terms (dark gray) and fermion-sfermion mixing matrices (light
    gray).
\label{fig:DipDistrib}}} }
\end{center}\end{figure}

\begin{figure}[t]\setlength{\unitlength}{1cm}
\begin{center}
  \raisebox{4cm}{
\parbox[b]{5cm}{\caption{Scatter plot for $d_N/d_e$ as generated from
    complex soft terms ($\bullet$) and fermion-sfermion mixing matrices
    ($\circ$) as function of $m_{\tilde{e}_R}/M_2$ and for random acceptable
    spectra.
\label{fig:dn/de}}} }
\begin{picture}(10,9.5)
  \put(0,0){\includegraphics{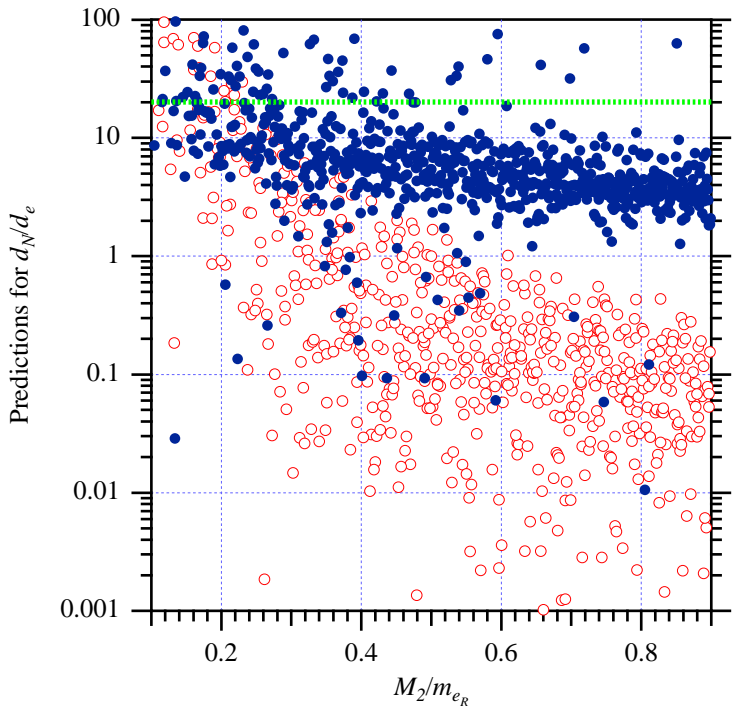}}
\end{picture}
\end{center}\end{figure}

We can now make a numerical computation, in the space of the parameters, of
$d_e$ and $d_N$ in the two cases.  As we have said, most important is the
correlation between $d_e$ and $d_N$. For this reason, the results are presented
as a scatter plot in the plane $(d_e,d_N)$ in fig.s~\ref{fig:dnde}. As
mentioned, the ``complex soft terms'' case depends on two phases $\phi_A$ and
$\phi_B$, whereas the ``unified theory'' case considered here depends on one
combination of phases only, $\phi$, which includes the standard CKM phase
entering the normal charged-current weak-interaction vertex.  In
fig.~\ref{fig:dnde}a we have taken a uniform random distribution of $\phi_A$,
$\phi_B$ and $\phi$ between $0$ and $2\pi$.  In fig.~\ref{fig:dnde}b the CP
violating phases are uniformly distributed in logarithmic scale, with the two
soft-term phases, $\phi_A$ and $\phi_B$, kept within the same order of
magnitude.  The other parameters are made to vary in such a way that
$$45\GeV<m_{\tilde{e}_R}<500\GeV,\qquad -3<{A_e \over m_{\tilde{e}_R}}<3,\qquad
1.5<\tan\beta<5,\qquad 0< M_2 <m_{\tilde{e}_R}.$$ This easily covers the range
of values for the relevant sparticle masses explorable at LHC by direct pair
production. Both squarks and gluinos go above $1\TeV$. We have checked that the
results do not change in any significant way if the sampling of the parameters
is uniformly distributed at the level of the initial conditions at the large
scale or at the level of the ``low energy'' parameters.

For the EDMs characteristic of the supersymmetric unified theory, the Yukawa
coupling of the top quark at the unification scale, $\lambda_{t\rm G}$, plays a
crucial role~\cite{8Art}. In fig.s~\ref{fig:dnde}, $\lambda_{t\rm G}$ is taken
to vary between $0.5$ and $1.4$. From extrapolation of the top Yukawa coupling
in the ``low energy'' range, we know that $\lambda_{t\rm G}$ should be bigger
than $0.5\div 0.6$ and that its preferred value from bottom-tau unification is
above unity. For values greater than one, $\lambda_{t\rm G}$ rapidly reaches an
infrared fixed point value in its behavior from $M_{\rm Pl}$ to $M_{\rm G}$:
$1.36$ is such a value for an SO(10) $\beta$-function coefficient of
$-3$~\cite{8Art}.

\section{}
Two facts are apparent from fig.s~\ref{fig:dnde}.  If the various phases are
not constrained to be small, the EDMs generated by the complex soft terms are
generally not consistent with the present bounds, unlike the case for the
effect in the unified theories. Most important for the purpose of this paper is
the rather significant correlation that exists between $d_e$ and $d_N$ when
they are generated by the phases in supersymmetry breaking terms: {\em the
  ratio of $d_N$ over $d_e$ is most often between 2 and 10 and is generally
  significantly bigger than in the unified theory case\/} irrespective of the
values of the CP violating phases (see the distributions of $d_N/d_e|_{\rm
  GUT}$ and of $d_N/d_e|_{\rm soft~terms}$ in fig.~\ref{fig:DipDistrib}).  The
ratio between the present upper bounds on $d_N$ and $d_e$ is
$d_N^{1995}/d_e^{1995}\approx 20$.

The qualitative reasons for the different behavior of the ratio $d_N/d_e$ in
the two cases are the following.

The lower correlation between $d_N$ and $d_e$ in the ``grand unified'' case,
apparent from fig.s~\ref{fig:dnde}, is an effect of the top Yukawa coupling,
which affects the low energy parameters related to the third generation in an
important way, and amplifies their dependence on the initial conditions.  One
rests here on the assumption of universality for the sfermion masses at the
large scale, that will have to be eventually checked by direct mass
measurements.

The relative value of $d_N/d_e|_{\rm SO(10)}$ versus $d_N/d_e|_{\rm
  soft~terms}$ has a definite pattern.  The EDMs generated by the complex soft
terms are proportional to the light fermion masses, unlike the case for the
unified theory, so that
\begin{equation}
  \frac{d_N/d_e|_{\rm soft~terms}}{d_N/d_e|_{\rm SO(10)}~~~} \simeq
  \frac{d_d/d_e|_{\rm soft~terms}}{d_d/d_e|_{\rm SO(10)}~~~}\propto
  \frac{m_d/m_e}{m_b/m_\tau } \simeq 5\pm2.
\end{equation}
Furthermore, in the unified theory case, the GIM-like cancellation is more
effective in suppressing the quark DMs, relative to the electron EDM, because
of a gluino focussing effect~\cite{9Art}: the strong radiative corrections to
the squark masses, proportional to the gluino mass, are family independent and,
as such, counteract the splitting effect induced by the top Yukawa coupling.
This is apparent in fig.~\ref{fig:dn/de} where we show a scatter plot of the
ratio $d_N/d_e$ versus the ratio $M_2/m_{\tilde{e}_R}$ between the $\SU(2)_L$
gaugino mass term $M_2$ and the right-handed selectron mass, both computed at
the Fermi scale.  The gluino mass is $M_3= 3.6 M_2$. In this plot all the
points correspond to values of both $d_e$ and $d_N$ not excluded by the present
bounds.  Fig.~\ref{fig:dn/de} shows that $d_N/d_e|_{\rm SO(10)}$ may in fact
overlap or even exceed $d_N/d_e|_{\rm soft~terms}$ with a significant
probability {\em only\/} for relatively light gluinos. This could of course
also be another handle to try to distinguish the physical origin of the effects
that might be observed.

Finally, as far as the strong CP-violation source is concerned, it is clear
from eq.~(\ref{sys:dipoles}), (\ref{eq:limiti}) that it could only show up in a
$d_N$ signal but not in $d_e$.

To conclude, we think to have shown that the combined efforts of several
experiments under way to try to measure an EDM have a significant chance not
only of producing a positive signal but also of providing crucial information
to understand the physical origin of the signal itself. This supports our view
that the EDM experiments are among the few crucial experiments, that can be
conceived at all, able to provide evidence for supersymmetric unification.

\subsubsection*{Acknowledgements}
We thank L.J. Hall for having raised the question addressed in this paper.

\end{document}